\documentclass[a4paper,superscriptaddress,showkeys,prX,showpacs,twocolumn]{revtex4-1}
\usepackage{amssymb, amsmath, amsfonts,bm} 
\usepackage{physics} 
\usepackage{siunitx} 
\usepackage{IEEEtrantools} 
\usepackage{graphics,xcolor} 
\usepackage{epsfig}
\usepackage{pst-all}
\usepackage{hyperref} 
\usepackage[mathlines,running,modulo]{lineno}
\usepackage{relsize}
\usepackage[utf8]{inputenc}
\usepackage[toc]{appendix}

\newcommand{\varg}{\textsl{g}}
\newcommand{\ud}{\,\mathrm{d}}

\usepackage{dcolumn}
\usepackage{latexsym} 
\usepackage[version=3]{mhchem}

\begin{document}


\title{Unveiling Antiferromagnetic Resonance: A Comprehensive Analysis via the Self-Consistent Harmonic Approximation}

\author{G. C. Villela}
\email{gabriel.villela@ufv.br}
\affiliation{Departamento de Física, Universidade Federal de Viçosa, 36570-900, Viçosa, Minas Gerais, Brazil}

\author{A. R. Moura}
\email{antoniormoura@ufv.br}
\affiliation{Departamento de Física, Universidade Federal de Viçosa, 36570-900, Viçosa, Minas Gerais, Brazil}

\date{\today}

\begin{abstract}
The Self-Consistent Harmonic Approximation (SCHA) has demonstrated efficacy in discerning phase transitions and, more recently, 
in elucidating coherent phenomena within ferromagnetic systems. However, a notable gap in understanding arises when extending this 
framework to antiferromagnetic models. In this investigation, we employ the SCHA formalism to conduct an in-depth exploration of the 
Antiferromagnetic Resonance (AFMR) within both Antiferromagnetic (AF) and Spin-Flop (SF) phases. Our analysis includes 
thermodynamic considerations from both semiclassical and quantum perspectives, with comparisons drawn against contemporary experimental 
and theoretical data. By incorporating a treatment utilizing coherent states, we investigate the dynamics of magnetization precession, 
a fundamental aspect in comprehending various spintronic experiments. Notably, the SCHA demonstrates excellent agreement with existing 
literature, showcasing its simplicity and efficiency in describing AFMR characteristics, even close to the transition temperature.
\end{abstract}


\maketitle

\section{Introduction and motivation}
\label{sec.introduction}
The Condensed Matter Physics community faces numerous challenges, and the development of spintronic 
devices occupies a central place\cite{science294.1488,jmmm509.166711}. In order to replace electronic-based 
devices with those that use spin as a degree of freedom, it is essential to have the ability to 
manipulate spin currents. Magnetic insulators are particularly interesting for sustaining spin 
currents as they reduce energy losses and provide higher frequency operation than traditional electronic 
devices \cite{pr885.1}. In this scenario, Spin-Transfer Torque (STT) and Spin Pumping (SP) are often used 
for the creation, manipulation, and detection of spin currents in the vicinity of interfaces 
involving magnetic insulators. In the STT process, a spin current is injected into the insulator due to the 
spin accumulation near the interface\cite{jmmm159.L1,prb54.9353}. In contrast, SP involves the spin current 
generation by using an oscillating microwave field that provides an angular momentum leaking to the material in 
contact with the magnetic insulator\cite{prl88.117601}. Although these processes were initially described for Ferromagnetic 
(FM) samples, they work equally well in Antiferromagnetic (AFM) insulators\cite{prl113.057601}. Indeed, for 
many decades, a considerable fraction of spintronics was primarily based on FM with minor interest in AFM. However, 
more recently, AFM spintronics has gained attention and proved advantageous over FM applications\cite{natphy14.220,am2024.1521}. 
For instance, AFM insulators are insensitive to external magnetic perturbations and provide vanishing stray fields due to 
the absence of macroscopic magnetization. Additionally, due to the AFM coupling, AFM frequencies reach up to THz while FM 
frequencies are restricted to the order of GHz. A comprehensive review of AFM spintronics can be found in 
references \cite{ptrsa369.3098,natnano.3.231,rmp90.015005,rezende}.

\begin{figure}[h]
	\label{fig.vectorscha}
	\centering 
	\epsfig{file=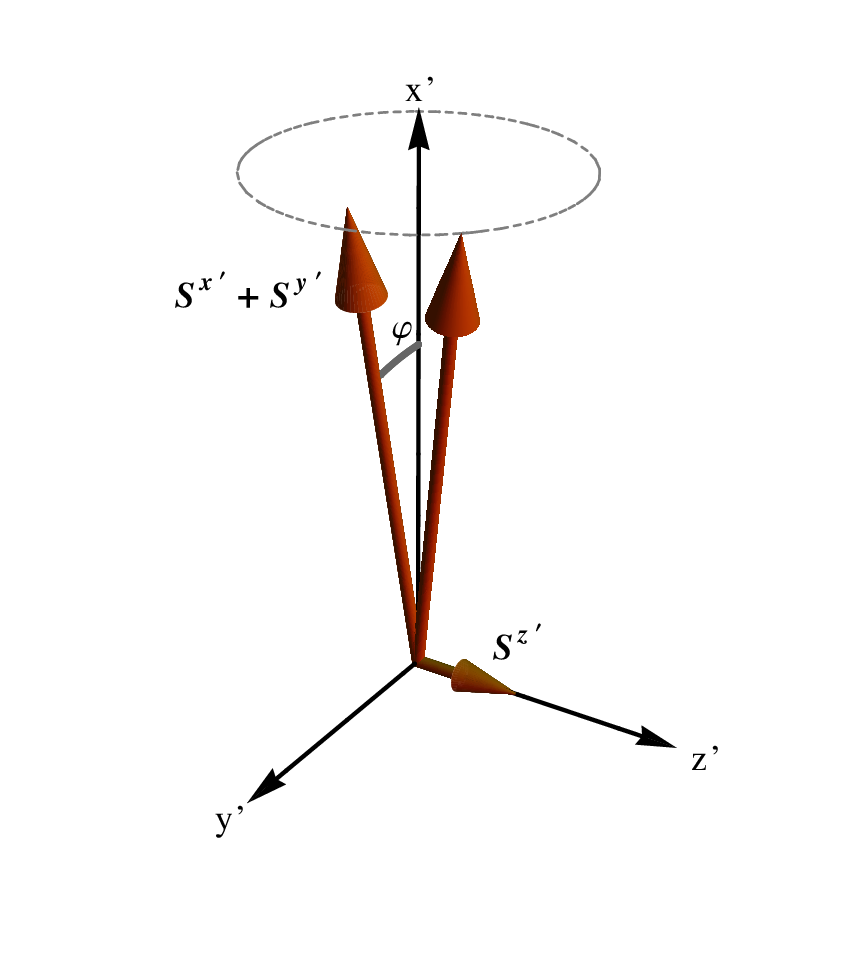,width=0.9\linewidth}
	\caption{The magnetization and the static magnetic field are defined along the $x$-axis. 
	The angle $\varphi$ is defined as the angle between the spin projection on the xy plane and the $x$-axis.}
\end{figure}

The SP mechanism is widely used to generate spin currents. It involves applying magnetic fields to create a 
coherent precession of magnetization. A static magnetic field is used to align the spin field, while an oscillating field 
supports the coherent dynamics. Then, by adjusting the static magnetic field intensity to give long-wavelength magnons with 
the same frequency as the oscillating field, a resonating condition occurs. This leads to an increasing population of low-energy 
magnons, which leak as spin currents into materials in contact with the magnet. During those magnetic resonance experiments, 
the entire spin field exhibits synchronous dynamics, and its behavior is formally described by coherent states, 
which were initially used to derive a fully quantum model of the radiation fields \cite{pr131.2766}, as well as coherent 
magnons \cite{pla29.47,pla29.616,prb4.201}. Provided that coherent states are the states that show minimum uncertainty, 
they are considered as the more classical-quantum states \cite{rmp62.867}. Consider, for instance, a particle in a harmonic 
potential represented by a Coherent State (CS). In this scenario, $\Delta x\Delta p=\hbar/2$, while the wave function describes a dispersionless 
wave packet that moves harmonically around the minimum of the potential. The resonating spin field exhibits similar semi-classical 
behavior, where the classical fields are represented by the phase angle $\varphi$ around the $z$-axis and the conjugate momentum 
associated, namely $S^z$. While in some cases, we adopt $S^z$ along the magnetization direction, it is not necessary, and 
we conveniently chose $S^z$ and $S^y$ as transverse components, while the magnetization is defined along the $x$-axis, as shown 
in Fig. \ref{fig.vectorscha}. Here, we are adopting $\varphi \ll 1$, and thus, the transverse spin components $S^y$ and $S^z$ are 
much smaller than the longitudinal component $S^x$. In addition, provided that $S^z\propto \dot{\varphi}$, both fields $S^y$ and $S^z$ 
show an oscillating behavior during the magnetization precession. From the classical perspective, the fields $\varphi_i$ and $S^z_j$, on 
sites $i$ and $j$, respectively, satisfy the Poisson bracket $\{\varphi_i,S^z_j\} = \delta_{ij}$, and the quantization is achieved by 
promoting the fields to operators that obey the commutation relation $[\varphi_i,S^z_j] = \delta_{ij}$. Similar to the particle
case, the operators obey the local equality of minimum uncertainty $\Delta\varphi\Delta S^z=1/2$, which justifies the semiclassical 
magnetization behavior of the spin. Therefore it is natural to adopt $\varphi$ and $S^z$ as the fundamental operators for 
describing resonance of magnetic models and the SCHA is the more convenient representation to be used.

The fundamental concept underlying the SCHA formalism involves substituting the original spin Hamiltonian with an 
alternative formulation characterized by solely quadratic terms in the canonically conjugate fields (or operators, within the 
quantum framework) $\varphi$ and $S^z$. Notably, the SCHA diverges from the Holstein–Primakoff (HP) formalism by incorporating spin 
fluctuations through the self-consistent determination of renormalization parameters, while the latter considers spin-wave 
interaction by including terms of quartic-order, or higher, in the Hamiltonian. Consequently, the SCHA model exhibits a dual nature, 
being both straightforward and precise in elucidating the thermodynamics of ordered phases in magnetic materials. Indeed, over 
the years, the SCHA has been successfully used to determine the critical temperature \cite{prb49.9663,pla202.309,prb51.16413,prb54.3019,ssc104.771,prb59.6229}, the topological BKT 
transition \cite{pla166.330,prb48.12698,prb49.9663,prb50.9592,ssc100.791,prb53.235,prb54.3019,prb54.6081,ssc112.705,epjb2.169,pssb.242.2138,prb78.212408,jmmm452.315}, 
and the large-D quantum phase transition \cite{pasma373.387,jpcm20.015208,pasma388.21,pasma388.3779,jmmm357.45} 
in a wide variety of magnetic models. Furthermore, the SCHA provides a highly convenient formalism for 
describing coherent states in ferromagnetic models and spintronic experiments, as exemplified in the 
references \cite{jmmm472.1} and \cite{prb106.054313}.

In the present study, we extend the preceding SCHA outcomes, originally derived for ferromagnetic models, to systematically examine 
Antiferromagnetic Resonance (AFMR). While the fundamental conceptualization of AFMR parallels that of the ferromagnetic model, 
nuanced considerations demand scrutiny. For instance, the presence of diverse phases, contingent upon the static magnetic field 
intensity, is well-established, presenting a distinctive characteristic absent in Ferromagnetic Resonance (FMR). Consequently, 
through the application of the SCHA formalism, we obtain a comprehensive elucidation of the coherent behavior of Antiferromagnetic models. 
This detailed analysis represents a crucial advancement in characterizing the intricacies of antiferromagnetic spintronics.

\section{Brief Review of AF models}
\label{sec.review}

To begin with, we first briefly review the basic concepts of AFMR. We adopt a simple AF model given by the following Hamiltonian
\begin{equation}
\label{eq.hamiltonian}
	H = 2J \sum_{\langle i,j \rangle} \bm{S}_i^\prime \cdot \bm{S}_j^\prime - D \sum_i \left( S_{i}^{x\prime} \right)^2 - \varg\mu_B \sum_i \bm{B}_i^\prime (t)\cdot\bm{S}_i^\prime,
\end{equation}
where $J>0$ is the AF exchange coupling constant, $\langle i,j\rangle$ represents nearest-neighbor spin interactions, $D>0$ 
is a single-ion anisotropic constant, and 
$\bm{B}^\prime(t) = B_x^\prime \hat{\imath}^\prime + B_y^\prime(t) \hat{\jmath}^\prime + B_z^\prime(t) \hat{k}^\prime$ 
is the magnetic field. We are adopting prime notation to represent spin components in the laboratory reference frame, while 
the notation without prime will be used to designate the local reference frame, as described below. Additionally, to clarify 
the notation, the direction index will be write as either a subscript or superscript index, with no physical difference between them. 

For the sake of clarity and without affecting the outcomes, we have opted to align the spin 
parallel to the magnetic moment in order to simplify the analysis. Additionally, 
for convenience, we define the $x$-axis, the longitudinal direction, along the 
anisotropy axis while the transverse components lay on the $yz$-plane. 
The static magnetic field $B_x^\prime$, which determines the kind of phase presented by the model, is applied parallel to
the anisotropic axis while $B_y^\prime(t)$ and $B_z^\prime(t)$ are the transverse magnetic field components. 
Therefore, we are leading with an easy-axis model; however, 
including easy-plane anisotropy is straightforward and requires minor changes in the next developments. Since there is no 
meaningful magnetization, and so the demagnetizing field vanishes, we write $\bm{B}^\prime=\mu_0 \bm{H}_\textrm{ext}^\prime$, 
where $\bm{H}_\textrm{ext}^\prime$ is the external H-field, which is regulated in laboratory. In addition, due to the small 
intensity of the transverse component, $B_y^\prime\approx B_z^\prime\ll B_x^\prime$, we can consider the 
oscillating field as a perturbation. 

\begin{figure}[h]
	\label{fig.phases}
	\centering 
	\epsfig{file=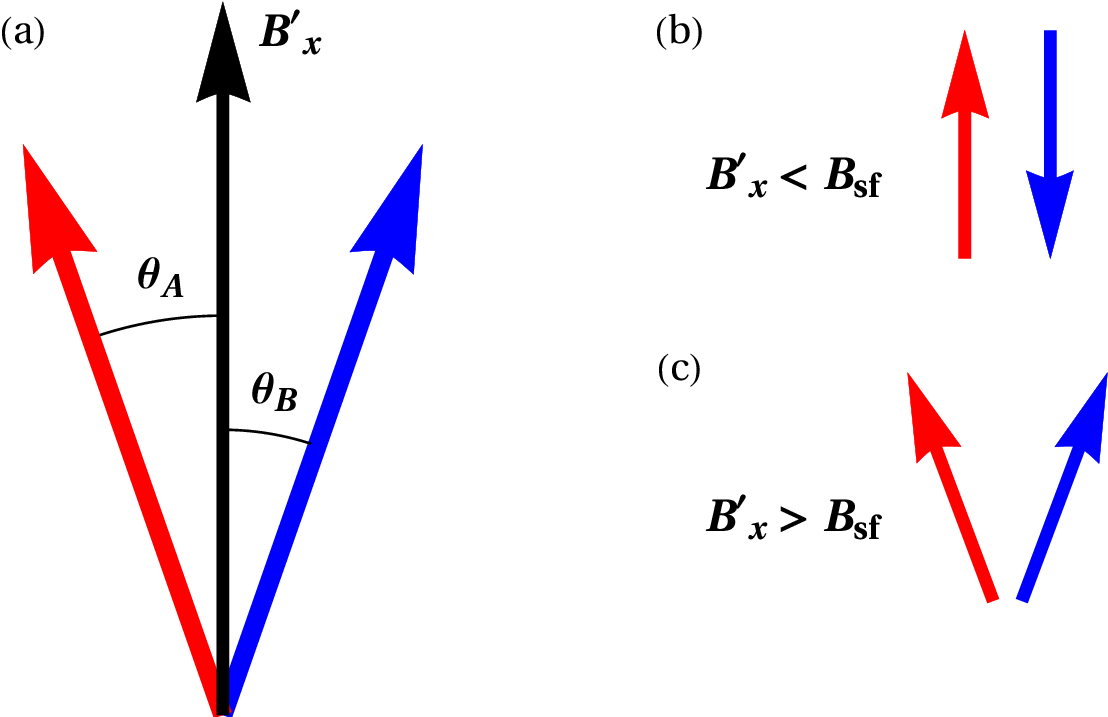,width=0.75\linewidth}
	\caption{(color online) (a) A general representation of the sublattice magnetizations. The red vector stands for the sublattice A, 
    and the blue represents the sublattice B. (b) For $B_x<B_\textrm{sf}$, we have the AF phase, which spin field solution is given 
    by $\theta_A=0$ and $\theta_B=\pi$. (c) For $B_x^\prime>B_\textrm{sf}$, the energy of the SF phase is minimized by the angles
    $\theta_A=\theta_B=\cos^{-1}[B_x^\prime/(2B_E-B_D)]$.}
\end{figure}

Depending on the intensity of $B_x^\prime$, we can get two different phases resulting from the energetic dispute between the Exchange and 
Zeeman energies. For small $B_x^\prime$ field, the Exchange energy prevails and the spin field assumes the usual AF phase (N\'{e}el
ordering) where one sublattice magnetization points in the opposite direction of the other one. In the strong field limit, the Zeeman
energy exceeds the Exchange term and the total energy is minimized by the so-called Spin-Flop (SF) phase, where the sublattice
magnetization assume the configuration shown in Fig. \ref{fig.phases} with $\theta_A=-\theta_B$. To determine the critical field $B_\textrm{sf}$,
where $sf$ stands for spin-flip, that separates the two phases, we consider the uniform sublattice magnetization limit defined by 
$\bm{M}_l=\gamma\hbar\bm{S}_l/2V_c$, where $\gamma=g\mu_B/\hbar=2\pi\times28$GHz/T is the gyromagnetic ratio, 
$V_c$ is the unit cell volume, and $l=A,B$ indicates the magnetization sublattice. Disregarding the small oscillating magnetic 
field, we obtain an energy expression depending on the local angles given by
\begin{IEEEeqnarray}{C}
	E(\theta_A,\theta_B)=\frac{\gamma\hbar \mathcal{N}S}{2}\left[B_E \cos(\theta_A+\theta_B)-\frac{B_D}{2}(\cos^2\theta_A+\right.\nonumber\\
	\left.+\cos^2\theta_B)-B_x^\prime(\cos\theta_A+\cos\theta_B)\right],
\end{IEEEeqnarray}
where $\mathcal{N}$ is the number of spins and the exchange and anisotropic fields are 
defined by $B_E=2zJS/\gamma\hbar$ and $B_D=2DS/\gamma\hbar$, respectively. The energy 
minimization provides the solutions $\theta_A=0$ and $\theta_B=\pi$ (AF phase) or $\theta_A=-\theta_B=\arccos[B_x^\prime/(2B_E-B_D)]$ (SF phase), 
and the analysis of the energy shows that the critical field is given by $B_\textrm{sf}=\sqrt{2B_EB_D-B_D^2}$. Therefore, as the field intensity
increases from small values, the N\'{e}el ordering decreases until the system undergoes a first-order phase transition at $B_x^\prime=B_\textrm{sf}$.

Under the constraint of a uniform spin field, corresponding to the $q=0$ magnons, the macroscopic magnetization exhibits behavior 
akin to a magnetic dipole undergoing precession within the influence of the static field $B_x^\prime$. Consequently, the dynamic evolution 
is governed by the Landau-Lifshitz (LL) equation, expressed as $d \bm{M}_l/dt=\gamma \bm{M}_l\times \bm{B}_{l,\textrm{eff}}$, 
where $\bm{B}_{l,\textrm{eff}}=-\partial H/\partial \bm{M}_l$ represents the effective magnetic field acting on the sublattice $l$. 
Substituting the magnetization ansatz $\bm{M}_l(t)=M_l^x\hat{\imath}+(m_l^y\hat{\jmath}+m_l^z\hat{k})e^{-i\omega_0 t}$ into the LL equation, 
we obtain a coupled set of ODEs that provides the $q=0$ magnon frequencies $\omega_0=\gamma B_x^\prime\pm \gamma (2 B_E B_D +B_D^2)^{1/2}$. In the 
absence of an external magnetic field ($B_x^\prime=0$), two degenerate modes emerge, precessing in opposite orientations. 
The frequency of the clockwise mode decreases with increasing $B_x^\prime$ and vanishes at the critical value 
$(2 B_E B_D +B_D^2)^{1/2}$, which is close to $B_\textrm{sf}$ for small anisotropies.

As usual, to determine the magnon spectrum of Eq. (\ref{eq.hamiltonian}), it is advantageous to transform the spins into a local reference 
frame, thereby establishing a collinear spin field. Considering a generic scenario, wherein each sublattice manifests a distinct rotating angle, 
a rotation around the $z$-axis is implemented to attain the new spin fields, in the local reference frame,
\begin{IEEEeqnarray}{rCl}
\label{eq.rotation}
	S_{r,l}^{x\prime} &=& S_{r,l}^x \cos\theta_l - S_{r,l}^y \sin\theta_l \nonumber\\
	S_{r,l}^{y\prime} &=& S_{r,l}^x \sin\theta_l + S_{r,l}^y \cos\theta_l \nonumber\\
	S_{r,l}^{z\prime} &=& S_{r,l}^z,
\end{IEEEeqnarray}
where $r$ represents the AF unit cell position and $l=A,B$ is the sublattice index. To avoid confusion, we reserve
the $i,j$ indexes for site position, which is  equivalently represented by the index pair $r,l$. 

\section{SCHA}
\label{sec.SCHA}

\subsection{Semiclassical approach}
\label{subsec.semiclassical}
In the preceding section, we derived the magnon energy for $q=0$ utilizing the LL phenomenological approach. 
To determine the complete energy spectrum, various established methodologies can be employed, including the HP
spin representation. While the HP formalism yields good results in scenarios involving non-interacting spin-waves, its 
extension to the high magnon density necessitates the inclusion of quartic-order or higher terms. 
Assessing the impacts of these high-order interaction terms is a complex undertaking, 
and whenever possible, it is advisable to avoid it. Conversely, the SCHA  
presents a quadratic model that yields satisfactory outcomes even close to the phase transition temperature. 
This agreement is achieved by the temperature-dependent renormalization parameters, which consider magnon interactions 
beyond the scope of second-order expansion. In this context, we employ the SCHA to explore the model 
described by Equation (\ref{eq.hamiltonian}). Initially, we adopt the semiclassical limit to express the spin components in 
terms of the fields $\varphi$ and $S^z$. Subsequently, quantization is obtained by promoting the field to appropriate operators.

From the semiclassical perspective, the spin field is written in terms of the canonically conjugate fields $\varphi$ 
and $S^z$ as $\bm{S}=\sqrt{S^2-(S^z)^2}(\cos\varphi\hat{\imath}+\sin\varphi\hat{\jmath})+S^z\hat{k}$, which provides 
the Hamiltonian (in the local reference frame)
\begin{IEEEeqnarray}{C}
	\label{eq.Hfull}
	H=2J\sum_{\langle i,j\rangle}[f_if_j(\cos\Delta\theta\cos\Delta\varphi_{ij}+\sin\Delta\theta\sin\Delta\varphi_{ij})+\nonumber\\
	+S_i^z S_j^z]-\sum_i\left\{\frac{f_i^2}{2}D(\cos 2\theta_i\cos 2\varphi_i-\sin 2\theta_i \sin 2\varphi_i\right.+\nonumber\\
	+1)\left.+g\mu_B[f_i (B_i^x\cos\varphi_i+B_i^y\sin\varphi_i)+ B_i^zS_i^z]\right\},
\end{IEEEeqnarray}
where $f_i=\sqrt{S^2-(S_i^z)^2}$, $\Delta\theta=\theta_A-\theta_B$, $\Delta\varphi_{ij}=\varphi_j-\varphi_i$,
$B_i^x=B^{x\prime}\cos\theta_l+B_i^{y\prime}\sin\theta_i$, $B_i^y=B_i^{y\prime}\cos\theta_i-B^{x\prime}\sin\theta_i$, and
$B_i^z=B_i^{z\prime}$. We consider that the static uniform magnetic field is uniform with $B^x\gg B_i^y,B_i^z$, and
$\theta_i$ can be $\theta_A$ or $\theta_B$ depending on the sublattice site. For determining the quadratic model, we expand 
the Hamiltonian up to second order in $\varphi$ and $S^z$. In a naive approximation, the expansion can be done without 
any correction. However, better results are obtained with the inclusion of 
renormalization parameters that consider the contributions of higher-order terms. In the SCHA, we include a renormalization factor $\rho$ for
each term that presents a trigonometric function expansion. Therefore, in the series expansion, we replace $\varphi$ by 
$\sqrt{\rho}\varphi$, in which the renormalization parameter is then found by solving a self-consistent equation. After
a straightforward procedure, we write the Hamiltonian as $H=E_0+H_1+H_2$, where
\begin{IEEEeqnarray}{rCl}
	E_0&=&\gamma\hbar NS[B_E\cos\Delta\theta-\frac{B_D}{2}(\cos^2\theta_A+\cos^2\theta_B)-\nonumber\\
	&&-(B_A^x+B_B^x)] 
\end{IEEEeqnarray}
is the ground state energy, 
\begin{IEEEeqnarray}{l}
\label{eq.H1}
	H_1=\gamma\hbar\sum_q\left\{S\sqrt{\rho}\left[\sqrt{N}(B_E\sin\Delta\theta+\frac{B_D}{2}\sin 2\theta_A)\delta_{q,0}-\right.\right.\nonumber\\
	\left.-B_{A,q}^y\right]\varphi_{A,q}+S\sqrt{\rho}\left[\sqrt{N}(-B_E\sin\Delta\theta+\frac{B_D}{2}\sin 2\theta_B)\delta_{q,0}-\right.\nonumber\\
	\left.\left.-B_{B,q}^y\right]\varphi_{B,q}-B_{z,q}(S_{A,q}^z+S_{B,q}^z)\right\}
\end{IEEEeqnarray}
is a linear term for uniform fields that will be considered as a potential, responsible for the coherent state, and
\begin{equation}
\label{eq.H2}	
	H_2=\frac{1}{2}\sum_q\sum_{ll^\prime}[S^2\rho\bar{\varphi}_l h_{ll^\prime}^\varphi\varphi_{l^\prime}+\bar{S}_l^z h_{ll^\prime}^z S_{l^\prime}^z]
\end{equation}
is the quadratic  Hamiltonian, which represents the non-interacting spin-waves endowed the temperature dependent renormalization
parameter $\rho$. Here, to make the notation clear, we hidden the momentum index in the $\varphi$ and $S^z$ field as well as
in the $h$ matrix coefficients. The momentum space is defined in relation to the reciprocal lattice of AF unit cells, 
and the summation over momenta is evaluated over the first Brillouin zone. $N$ specifies the number of unit cells, while
$\mathcal{N}=2N$ is the number of sites. In above equation, the matrix coefficients are given by
\begin{IEEEeqnarray}{rCl}
	\IEEEyesnumber
	\IEEEyessubnumber*
	h_{AA}^\varphi&=&\frac{\gamma\hbar}{S}(-B_E\cos\Delta\theta+B_D\cos 2\theta_A+B_A^x)\\
	h_{AB}^\varphi&=&h_{BA}^\varphi=\frac{\gamma\hbar}{S}B_E\cos\Delta\theta\gamma_q\\
	h_{BB}^\varphi&=&\frac{\gamma\hbar}{S}(-B_E\cos\Delta\theta+B_D\cos 2\theta_B+B_B^x)\\
	h_{AA}^z&=&\frac{\gamma\hbar}{S}(-B_E\cos\Delta\theta+B_D\cos^2\theta_A+B_A^x)\\
	h_{AB}^z&=&h_{BA}^z=\frac{\gamma\hbar}{S}B_E\gamma_q\\
	h_{BB}^z&=&\frac{\gamma\hbar}{S}(-B_E\cos\Delta\theta+B_D\cos^2\theta_B+B_B^x),
\end{IEEEeqnarray}
where $\gamma_q=z^{-1}\sum_{\delta r} e^{i\bm{q}\cdot\bm{\delta r}}$ is the structure factor of a lattice defined by $z$ 
nearest-neighbor spins located at $\bm{\delta r}$ positions. Note that, in the particular AF or SF cases, the coefficients
are simplified and the evaluation becomes easier. The renormalization parameter is determined from the
weighted average
\begin{IEEEeqnarray}{rCl}
	\label{eq.rho}
	\rho&=&\frac{1}{Z_\rho}[-2\cos\Delta\theta B_E\rho_E+(\cos 2\theta_A+\cos 2\theta_B)B_D\rho_D+\nonumber\\
	&&+(B_A^x+B_B^x)\rho_Z],
\end{IEEEeqnarray}
where $Z_\rho=(h_{AA}^\varphi+h_{BB}^\varphi)S/(\gamma\hbar)$, and the exchange, anisotropic and Zeeman 
renormalization parameters are given by the self-consistent equations
\begin{equation}
	\rho_E=\left(1-\frac{\langle (S^z)^2\rangle_G}{S^2}\right)\exp\left(-\frac{1}{2}\langle\Delta\varphi^2\rangle_G\right),
\end{equation}
\begin{equation}
	\rho_D=\left(1-\frac{\langle (S^z)^2\rangle_G}{S^2}\right)\exp\left(-2\langle\varphi^2\rangle_G\right),
\end{equation}
and 
\begin{equation}
	\rho_Z=\left(1-\frac{\langle (S^z)^2\rangle_G}{2S^2}\right)\exp\left(-\frac{1}{2}\langle\varphi^2\rangle_G\right),
\end{equation}
respectively. The notation $\langle\ldots\rangle_G$ represents Gaussian averages, which are evaluated using 
the quadratic Hamiltonian $H_2$. The demonstration of the self-consistent equation is given in Appendix (\ref{app.SCHA}).

The quantization of Hamiltonian (\ref{eq.H2}) provides the renormalized magnon spectrum. However, we can also determine the 
spin-wave energy through the field approach of the SCHA. As an illustrative example, let us consider the AF phase, characterized by $\theta_A=0$ 
and $\theta_B=\pi$.	In this case, $h_{AA}^\varphi=h_{AA}^z=h_{AA}$, $h_{BB}^\varphi=h_{BB}^z=h_{BB}$, 
and $-h_{AB}^\varphi=h_{AB}^z=h_{AB}$. In the laboratory reference frame, we assume that the spin field precess with 
time-independent $S^{x\prime}$ component, while non-localized fluctuations are expressed as
$S_q^{+\prime}(t)=S_q^{y \prime}(t)+iS_q^{z\prime}(t)=S_q^{+\prime}(0)e^{-i\omega_q^\alpha t}$. 
Employing a semiclassical analysis, the spin dynamics is then achieved from the Hamilton equations, yielding
\begin{equation}
\hbar\dot{\varphi}_l=\frac{\partial H_2}{\partial \bar{S}_l^z}=h_{ll}S_l^z+h_{ll^\prime}S_{l^\prime}^z,
\end{equation}
and
\begin{equation}
\hbar\dot{S}_l^z=-\frac{\partial H_2}{\partial\bar{\varphi}_l}=(-h_{ll}\varphi_l+h_{ll^\prime}\varphi_{l^\prime})S^2\rho,
\end{equation}
where $l=A,B$ and $l\neq l^\prime$. Since $S_q^{yl}\approx S\sqrt{\rho}\varphi_{l,q}$, it is easy to show that the non-trivial solution
of the dynamic equations results in the magnon frequency
\begin{equation}
\label{eq.omegasc}
\omega_q^\alpha=\sqrt{\rho}\gamma\left[ B_x^\prime+\sqrt{2B_E B_D+B_D^2+B_E^2(1-\gamma_q^2)}\right].
\end{equation}
Similarly, the magnon frequency ($\beta$-mode) is obtained considering $S_q^{+\prime}(t)=S_q^{+\prime}(0) e^{i\omega_q^\beta t}$.
In addition, note that, in the long wavelength limit, the SCHA results coincide with the uniform magnetization dynamics
obtained from the LL equation.

Note that $\sqrt{\rho}$ multiply each field term in Eq. ({\ref{eq.omegasc}). This leads to introduction of renormalized
fields denoted as $B_E^r=\sqrt{\rho}B_E$, $B_D^r=\sqrt{\rho}B_D$, and $B_x^{\prime r}=\sqrt{\rho}B_x^\prime$, with the
original values $B_E$, $B_D$, and $B_x^\prime$ representing the bare fields. At $T=0$, the semiclassical SCHA approach 
yields $\rho=1$ causing both renormalized and bare parameters to coincide. However, at finite temperature, 
the physically significant terms are the renormalized ones. In terms of the renormalized fields, the SCHA spin-wave energy 
is identical to the HP one. A comparable outcome arises from the quantum treatment.

\subsection{Quantum approach}
\label{subsec.quantum} 

Within the quantum framework, the fields $\varphi$ and $S^z$ are replaced by operators that satisfy the commutation relation 
$[\varphi_i,S_j^z]=\delta_{ij}$. Consequently, it becomes convenient to represent the Hamiltonian in terms of bosonic operators, 
which are defined by
\begin{IEEEeqnarray}{rCl}
	\label{eq.spinfieldA}
	\IEEEyesnumber
	\IEEEyessubnumber*
	\bar{\varphi}_{A,q}&=&\frac{1}{\sqrt{2}(S^2\rho)^{1/4}}(a_q^\dagger+a_{-q})\\
	\bar{S}_{A,q}^z&=&\frac{i(S^2\rho)^{1/4}}{\sqrt{2}}(a_q^\dagger-a_{-q})
\end{IEEEeqnarray}
for sublattice A, and
\begin{IEEEeqnarray}{rCl}
	\label{eq.spinfieldB}
	\IEEEyesnumber
	\IEEEyessubnumber*
	\bar{\varphi}_{B,q}&=&\frac{1}{\sqrt{2}(S^2\rho)^{1/4}}(b_q^\dagger+b_{-q})\\
	\bar{S}_{B,q}^z&=&\frac{i(S^2\rho)^{1/4}}{\sqrt{2}}(b_q^\dagger-b_{-q})
\end{IEEEeqnarray}
for sublattice B. It is straightforward to verify that the new operators satisfy the commutation relations 
$[a_q,a_q^\dagger]=1$, $[b_q,b_q^\dagger]=1$, and $[a_q,b_q]=0$. In addition, when we are dealing with small spin values in quantum 
regime, better results are obtained replacing $S$ by $\tilde{S}=\sqrt{S(S+1)}$. 

Adopting the bosonic operators, the Hamiltonian is then written as
\begin{equation}
H_2=\frac{1}{2}\sum_q X_q^\dagger \mathcal{H}_q X_q,
\end{equation} 
where we define the vector $X_q^\dagger=(a_q^\dagger\ b_q^\dagger\ a_{-q}\ b_{-q})$, and the matrix
\begin{equation}
\label{eq.matrixH}
\mathcal{H}_q=\left(\begin{array}{cccc}
	A_a & B_q & C_a & D_q \\
	B_q & A_b & D_q & C_b \\
	C_a & D_q & A_a & B_q \\
	D_q & C_b & B_q & A_b 
	\end{array}\right)=I\otimes \mathcal{H}_{1q}+\sigma_x\otimes \mathcal{H}_{2q},
\end{equation}
whose coefficients are given by
\begin{IEEEeqnarray}{C}
	A_a=\frac{\sqrt{\rho}\tilde{S}}{2}(h_{AA}^\varphi+h_{AA}^z), \ \ A_b=\frac{\sqrt{\rho}\tilde{S}}{2}(h_{BB}^\varphi+h_{BB}^z) \nonumber\\
	C_a=\frac{\sqrt{\rho}\tilde{S}}{2}(h_{AA}^\varphi-h_{AA}^z),\ \ C_b=\frac{\sqrt{\rho}\tilde{S}}{2}(h_{BB}^\varphi-h_{BB}^z) \nonumber\\
	B_q=\frac{\sqrt{\rho}\tilde{S}}{2}(h_{AB}^\varphi+h_{AB}^z), \ \	D_q=\frac{\sqrt{\rho}\tilde{S}}{2}(h_{AB}^\varphi-h_{AB}^z). \nonumber
\end{IEEEeqnarray}
As usual, we diagonalize $H$ through a Bogoliubov transformation on the bosonic operators\cite{pr139.a450}. 
This being so, we define new operators $\Psi_q^\dagger=(\alpha_q^\dagger\ \beta_q^\dagger\ \alpha_{-q}\ \beta_{-q})$ by the linear transformation
$X_q=T_q\Psi_q$. To ensure the bosonic relation commutation, the matrix transformation must obey the relation
$T_qGT_q^\dagger=G$, where $G=\sigma_z\otimes I$. Therefore,
\begin{equation}
	\label{eq.Hdiagonal}
	X_q^\dagger\mathcal{H}_q X_q=\Psi_q^\dagger G(T_q^{-1}G\mathcal{H}_qT_q)\Psi_q=\Psi_q^\dagger\Omega_q\Psi_q,
\end{equation}
where $\Omega_q$ is the diagonal matrix given by $\Omega_q=\textrm{diag}(\omega_q^\alpha,\omega_q^\beta,\omega_q^\alpha,\omega_q^\beta)$.
Technically, the matrix $T_q$ diagonalizes $G\mathcal{H}_q$; however, determining the transformation for the general case, as defined
by Eq. (\ref{eq.matrixH}), shows a highly intricate challenge. Fortunately, under certain symmetrical conditions, exact solutions can 
be easily derived for both the AF and SF cases.

Due to the transformation scenario, the more general transformation $T_q$ is written as
\begin{equation}
\label{eq.matrixT}
T_q=\left(\begin{array}{cc}
	U_q & V_q \\
	\bar{V}_q & \bar{U}_q
	\end{array}\right),
\end{equation}
where $U_q=[u_{mn}]$ and $V_q=[v_{mn}]$ are two-dimensional square matrices. Using Eq. (\ref{eq.Hdiagonal}), we found
that the submatrices of $T_q$ obey
\begin{IEEEeqnarray}{rCl}
    \label{eq.UV}
	\IEEEyesnumber
	\IEEEyessubnumber*
	\mathcal{H}_{1q} U_q+\mathcal{H}_{2q}\bar{V}_q&=&U_q\Omega_q,\\
	\mathcal{H}_{1q} V_q+\mathcal{H}_{2q}\bar{U}_q&=&-V_q\Omega_q.
\end{IEEEeqnarray}
In terms of the $u_{mn}$ and $v_{mn}$ coefficients, the new operators are related to the old ones by
$a_q=u_{11}\alpha_q+u_{12}\beta_q+v_{11}\alpha_{-q}^\dagger+v_{12}\beta_{-q}^\dagger$, and
$b_q=u_{21}\alpha_q+u_{22}\beta_q+v_{21}\alpha_{-q}^\dagger+v_{22}\beta_{-q}^\dagger$. Note that the matrix 
coefficients are not all independent, as the commutation relations between $a$ and $b$ impose constraints on them. 
For instance, $[a_q,a_q^\dagger]=1$ implies $|u_{11}|^2+|u_{12}|^2-|v_{11}|^2-|v_{12}|^2=1$. Furthermore, depending on 
the symmetries involved, we can parametrize the coefficients to facilitate the diagonalization process.

\textbf{AF phase} - Taking into account that $\theta_A=0$ and $\theta_B=\pi$ in the AF phase, the coefficients $B_q$, $C_a$, and $C_b$ vanish.
Therefore, the new operators $\alpha_q$ and $\beta_q$ are given by the transformation $U_q=\cosh\Theta_q I$, and $V_q=\sinh\Theta_q \sigma_x$, which
results in
\begin{IEEEeqnarray}{rCl}
	\label{eq.abAF}
	\IEEEyesnumber
	\IEEEyessubnumber*
	a_q&=&\cosh\Theta_q\alpha_q+\sinh\Theta_q\beta_{-q}^\dagger\nonumber\\
	b_q&=&\cosh\Theta_q\beta_q+\sinh\Theta_q\alpha_{-q}^\dagger.
\end{IEEEeqnarray}

The angle $\Theta_q$ is properly chosen to eliminate the Hamiltonian off-diagonal terms, which is achieved by
\begin{equation}
	\label{eq.ThetaAF}
	\tanh2\Theta_q=-\frac{2D_q}{A_a+A_b}=\frac{B_E \gamma_q}{B_E+B_D}.
\end{equation}
After replacing Eq. (\ref{eq.ThetaAF}) in the Hamiltonian, we obtain the diagonal model
$H_{AF}=E_0^{AF}+\sum_q(\epsilon_q^\alpha \alpha_q^\dagger\alpha_q+\epsilon_q^\beta \beta_q^\dagger \beta_q)$, where
$E_0^{AF}=E_0+\sum_q(\hbar\omega_q/2)$ is a constant energy, 
\begin{IEEEeqnarray}{rCl}
	\label{eq.H1H2}
	\IEEEyesnumber
	\IEEEyessubnumber*
	\epsilon_q^\alpha&=&\gamma\hbar B_x^{\prime r}+\hbar\omega_q,\\
	\epsilon_q^\beta&=&-\gamma\hbar B_x^{\prime r}+\hbar\omega_q,
\end{IEEEeqnarray}
and
\begin{equation}
\omega_q=\gamma\sqrt{2B_E^r B_D^r+(B_D^r)^2+(B_E^r)^2(1-\gamma_q^2)}.
\end{equation}
For small anisotropy field, the long-wavelength limit results in the relativistic dispersion relation
$\omega_q=(2/z)^{1/2}\gamma B_E^r aq$, where $a$ is the lattice spacing.
Observe that, in the above equations, the fields correspond to the previously defined renormalized ones. 
This scenario parallels to the semiclassical approach. However, the renormalization parameter now includes 
quantum fluctuations, which leads to a significant reduction in $\rho$, even at $T=0$. 
To compensate for small value of $\rho$, we must employ bare parameters larger than those in the 
semiclassical scenario. Nevertheless, this poses no issue as the physically relevant parameters are 
always the renormalized ones, which exhibit similarity in both cases.
\begin{figure}[h]
\centering 
\epsfig{file=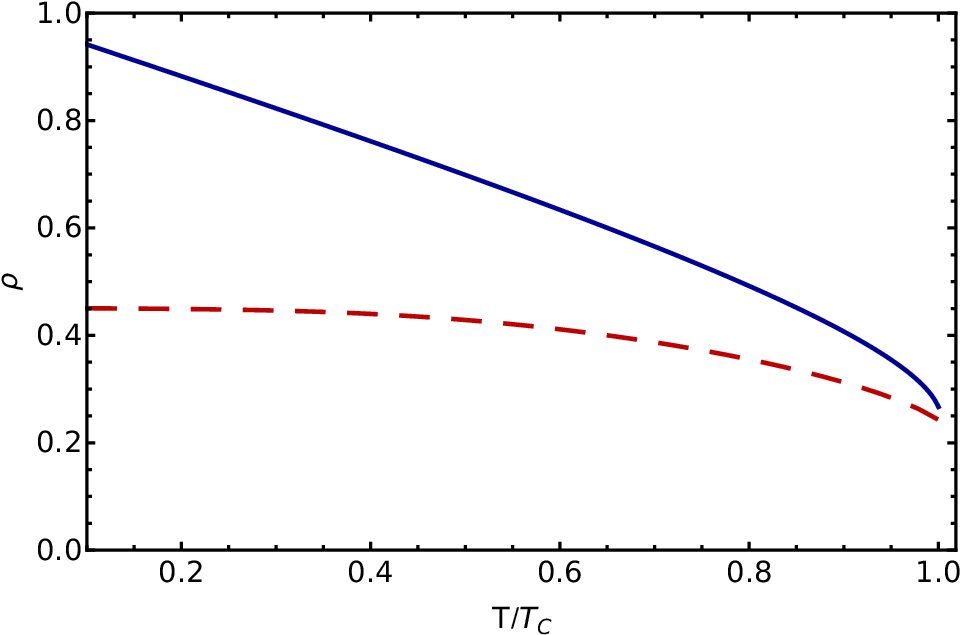,width=0.9\linewidth}
\caption{(color online) The renomalization parameter $\rho$ obtained from the semiclassical (solid blue curve) and 
quantum approach (dashed red curve) for the AF phase in the absence of external fields. 
Here, we use the values $J/k_B=1 K$, $D=0.1J$, $S=1$, and $z=6$.}
\label{fig.rhoAF}
\end{figure}

Fig. (\ref{fig.rhoAF}) plots the renormalization parameter as a function of the reduced temperature $T/T_c$, 
where $T_c$ denotes the temperature at which $\rho$ abruptly vanishes, indicating a non-physical behavior associated 
with a first-order transition in staggered magnetization. This anomaly is usually seen in theories relying 
on harmonic expansions. Despite this limitation, the SCHA method yields excellent results for temperatures 
$T<T_c$, with the formalism exhibiting issues only in close proximity to $T_c$.
It is noteworthy that, even in traditional spin representations like the HP formalism, accurately describing 
thermodynamics near the critical temperature proves challenging. Despite achieving excellent agreement in 
the low-temperature regime, when adopting the non-interacting spin-wave limit, the HP representation fails close
to $T_c$. In this case, it is necessary to consider the quartic-order interaction terms, which renormalize 
the spin-wave energy, for obtaining a more reasonable critical temperature \cite{pps82.992}. In contrast, the critical 
temperature obtained from the harmonic approximation closely aligns with the real critical temperature. 
Therefore, despite the imprecise behavior at $T=T_c$, $T_c$ can be adopted as a reasonable estimation for the transition temperature.

The energy results of the AF phase align with the semiclassical approach and are identical to the HP outcomes, considering 
that the renormalized fields replace the equivalent HP parameters. Additionally, the SCHA results give 
temperature-dependent corrections, which are disregarded in the linear spin-wave analysis.

\textbf{SF phase} - Considering $B_y^\prime,B_z^\prime\ll B_x^\prime$, as occurs in typical spintronic experiments, we obtain 
$h_{AA}^\varphi=h_{BB}^\varphi$ and $h_{AA}^z=h_{BB}^z$. Therefore, in the SF phase, $A_a=A_b=A=\gamma\hbar[-B_E^r\cos 2\theta+B_D^r(\cos^2\theta+\cos 2\theta)/2+B_x^{\prime r}\cos\theta]$, $B_q=\gamma\hbar B_E^r\gamma_q\cos^2\theta$, $C_a=C_b=C=-\gamma\hbar (B_D^r\sin^2\theta)/2$, and $D_q=-\gamma\hbar B_E^r\gamma_q \sin^2\theta$. Then, the
Hamiltonian is diagonalized by the transformations (chosen to provide positive angles in the long-wavelength limit)
\begin{equation}
\label{eq.matrixUsf}
U_q=\frac{1}{\sqrt{2}}\left(\begin{array}{cc}
	\cosh\Xi_q & \cosh\Phi_q \\
	-\cosh\Xi_q & \cosh\Phi_q
	\end{array}\right),
\end{equation}
and
\begin{equation}
\label{eq.matrixVsf}
V_q=\frac{1}{\sqrt{2}}\left(\begin{array}{cc}
	-\sinh\Xi_q & \sinh\Phi_q \\
	\sinh\Xi_q & \sinh\Phi_q
	\end{array}\right).
\end{equation}
The angles $\Xi_q$ and $\Phi_q$ are determined by solving Eq. (\ref{eq.UV}), which yields
\begin{IEEEeqnarray}{rCl}
	\IEEEyesnumber
	\IEEEyessubnumber*
	\tanh\Xi_q&=&\sqrt{\frac{A-B_q-\epsilon_q^\alpha}{A-B_q+\epsilon_q^\alpha}}\\
	\tanh\Phi_q&=&\sqrt{\frac{A+B_q-\epsilon_q^\beta}{A+B_q+\epsilon_q^\beta}},
\end{IEEEeqnarray}
with the respective magnon energies
\begin{IEEEeqnarray}{rCl}
	\IEEEyesnumber
	\IEEEyessubnumber*
	\epsilon_q^\alpha&=&\sqrt{(A-B_q)^2-(C-D_q)^2},\\
	\epsilon_q^\beta&=&\sqrt{(A+B_q)^2-(C+D_q)^2}.
\end{IEEEeqnarray}
Similarly to the AF case, the diagonal Hamiltonian in the SF phase is expressed as 
$H_{SF}=E_0^{SF}+\sum_q(\epsilon_q^\alpha \alpha_q^\dagger\alpha_q+\epsilon_q^\beta \beta_q^\dagger \beta_q)$,
where $E_0^{SF}=E_0+\sum_q(\epsilon_q^\alpha+\epsilon_q^\beta)/4$. After the SF transition, the 
alpha and beta mode show distinct behavior. The alpha mode exhibits a gapless relativistic energy spectrum, 
while the beta mode presents a gapped quadratic dispersion relation similar to the fluctuations of the 
ferromagnet model (in the presence of external magnetic field). 
When considering a small anisotropic field in the long-wavelength limit, we obtain $\epsilon_p^\alpha=pc$, 
with the spin-wave velocity $c=\sqrt{2/z}a\gamma B_E^r$, and $p=\hbar q$. For the beta mode, 
$\epsilon_p^\beta=\epsilon_0+p^2/2m$, where $\epsilon_0=\gamma\hbar\sqrt{(B_x^{\prime r})^2-(B_\textrm{sf}^r)^2}$ 
represents the rest energy, and
\begin{equation}
m=\frac{2z}{\gamma a}\frac{\sqrt{(B_x^{\prime r})^2-(B_\textrm{sf}^r)^2}}{4(B_E^r)^2-3(B_x^{\prime r})^2}
\end{equation}
performs the role of a mass term. Note that, at $B_x^{\prime r}=B_\textrm{sf}^r$, the mass vanishes and the beta mode
returns to the relativistic behavior (AF phase). The SCHA results coincide with the HP outcomes with the advantage of 
the temperature renormalization. The $\rho$ evaluation follows the previous procedure and the temperature 
behavior is similar to those obtained from AF phase.

\subsection{SCHA assessment}
\label{subsec:AFres} 
To assess the SCHA effectiveness in characterizing the thermodynamics of magnetic models, we employ the 
formalism to examine the properties of the antiferromagnets \ce{MnF_2}, \ce{FeF_2}, and \ce{RbMnF_3}. \ce{MnF_2} and
\ce{FeF_2} exhibit a body-centered tetragonal lattice and possess uniaxial anisotropy (interactions beyond 
the nearest neighbors, intra-sublattice, are very weak and hence disregarded in our calculations).
\ce{MnF_2} (spin $5/2$) has a small anisotropy resulting from dipolar interaction, which provides a N\'{e}el 
temperature of 66.5 K\cite{cjp88.771}. In \ce{FeF_2} (spin $S=2$), the anisotropy arises from orbital angular momentum that 
justifies the larger N\'{e}el temperature of 78.4 K\cite{jpcssp3.307}. The \ce{RbMnF_3} compound (spin $5/2$) shows 
a simple cubic structure, negligible anisotropy and N\'{e}el temperature of 83 K\cite{prb18.2346}. 
While the $J$ and $D$ values are widely known, we opt to determine the coupling constants
from experimental data to achieve a more precise alignment with the SCHA model.

Considering the free external magnetic field scenario, the magnon energy is expressed as
$\epsilon_q=\hbar\gamma B_E^r\sqrt{\Delta^2-\gamma_q^2}$, where $\Delta=1+B_D^r/B_E^r\gtrsim 1$. To
determine the renormalized exchange field $B_E^r$, we apply the method of least squares on the 
error $\sum_q(\epsilon_q-\epsilon_q^\textrm{exp})^2$, where $\epsilon_q^\textrm{exp}$ represent the
experimental energy data measure at $T_0$, and the momentum sum is computed over the experimental
values. The minimization with respect to $B_E^r$ provides
\begin{equation}
B_E^r=\frac{\sum_q\epsilon_q^\textrm{exp}\sqrt{\Delta^2-\gamma_q^2}}{\hbar\gamma\sum_q (\Delta^2-\gamma_q^2)},
\end{equation}
where we neglect the small contributions of $\Delta$ derivatives. In the initial step, we set $\Delta^{(0)}=1$ for
determining $B_E^r$, which is then used to obtain the next $\Delta^{(1)}=[1+(\epsilon_0^\textrm{exp}/\hbar\gamma B_E^r)^2]^{1/2}$. 
We iterate the process until achieving the $\Delta$ convergence. Solving the self-consistent equation at $T_0$, 
we obtain the renormalization $\rho(T_0)$, which results in the bare constant $B_E=B_E^r/\rho(T_0)$.

In all analyzed compounds, the quantum SCHA approach outperforms its semiclassical 
counterpart. For the \ce{MnF_2} at $T=10$ K, we determined the renormalized fields as $B_E^r=57.1$ T and $B_D^r=0.73$ T, 
resulting in a transition temperature of 65.9 K. For \ce{FeF_2}, the fields parameters (at $T=10$ K) are $B_E^r=57.2$ T and
$B_D^r=19.6$ T, while the N\'{e}el temperature is 78.6 K. Finally, for \ce{RbMnF_3} we found at $T=10$ K,
$B_E^r=77.0$ T and $B_D^r=0$, which yields the transition temperature of 83.9 K. 
Although the field constants are higher than those derived 
from the HP analysis (partially due to the disregarded intra-sublattice interactions), 
these values yield the optimal fit with experimental data and demonstrate excellent 
agreement with other thermodynamic outcomes. Table (\ref{tab.TN}) provides a summary of the temperatures obtained
from SCHA.
\begin{figure}[h]
\centering 
\epsfig{file=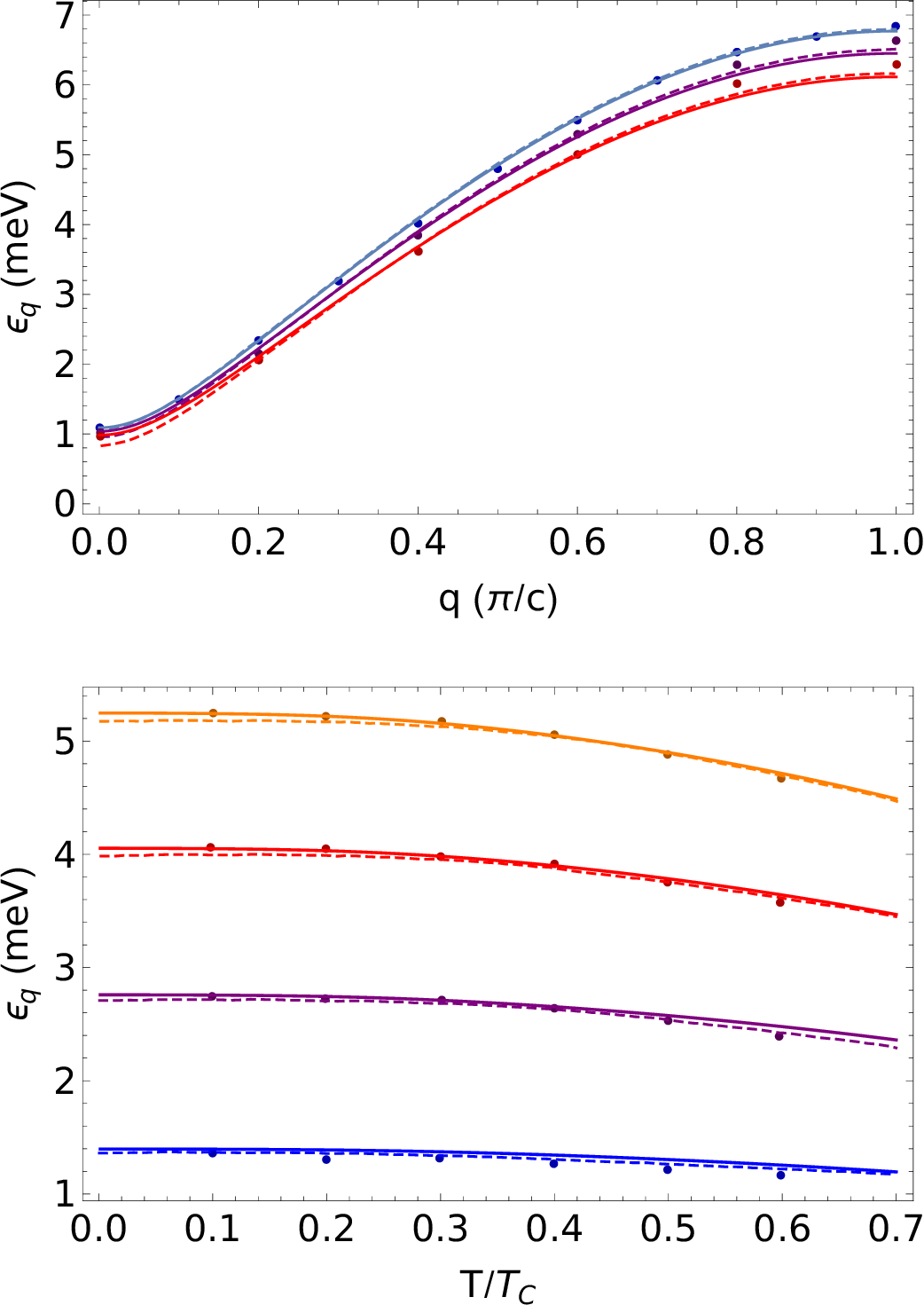,width=0.9\linewidth}
\caption{(color online) The renormalized energy spectrum obtained from SCHA (solid line) and HP formalism (dashed line) for
\ce{MnF_2} (above) and \ce{RbMnF_3} (below). In the \ce{MnF_2} plot, the wavenumber is measure along the $x$-direction and 
the curves indicate temperatures of 3K (red), 30 K (purple), and 40 K (blue). For the \ce{RbMnF_3}, the curves designate
$aq$ equals to $0.1\pi$ (blue), $0.2\pi$ (purple), $0.3\pi$ (red), and $0.4\pi$ (orange). 
The experimental values were extracted from Ref. \cite{science312.1926} (\ce{MnF_2}) and \cite{prb18.2346} (\ce{RbMnF_3}).}
\label{fig.MnF2}
\end{figure}

Fig. (\ref{fig.MnF2}) shows the renormalized spectrum energy obtained through SCHA, HP representation
and experimental data for \ce{MnF_2}\cite{science312.1926} and \ce{RbMnF_3}\cite{prb18.2346}. 
The HP results are performed by using Random-Phase Approximation for four-magnon 
interactions\cite{prb14.2939}. At each temperature, the HP renormalization is numerically evaluated through self-consistent
iterating. For SCHA, we determine the best field parameters using the reference temperature at 3 K, and renormalize 
the high-temperature field parameters through $\rho(T)$. The SCHA, which is more simple to apply, provides slightly superior 
outcomes compared to the HP formalism.

\begin{table}[h]
\label{tab.TN}
\begin{ruledtabular}
\begin{tabular}{cccc}
Material &SC-SCHA& Q-SCHA& Exp.\\
\colrule
\ce{MnF_2} & 49.3 K & 65.9 K & 67.0 K\footnote{Canadian J. Phys. 88, 771 (2010)}\\
\ce{FeF_2} & 54.5 K & 78.6 K & 78.4 K\footnote{J. Phys. C: Solid State Phys. 3, 307 (1970)}\\
\ce{RbMnF_3} & 63.6 K & 83.9 K & 83.0 K\footnote{Phys. Rev. B 18, 2346 (1978)}
\end{tabular}
\end{ruledtabular}
\caption{SCHA N\'{e}el temperatures for \ce{MnF_2}, \ce{FeF_2}, and \ce{RbMnF_3}
 obtained from the semiclassical (SC) and quantum (Q) approaches. The last column shows the 
 experimental data.}
\end{table}

\section{Magnon coherent states}
\label{sec.CS}
The static and dynamic components of the magnetic field $\bm{B}$ play a crucial role in facilitating AFMR-driven 
spin pumping. In a typical AFMR experiment, an alternating field, which we adopt perpendicular to the static 
field, at microwave frequencies induces the spin field to oscillate around the direction determined by the static field. 
While maintaining a constant frequency $\Omega$ for the oscillating field, adjustments are made to the static field to 
achieve the resonance condition of excited magnons, reached when the energy of the $q=0$ magnon mode is equal 
to $\hbar\Omega$. Upon successful attainment of the resonance 
condition, the entire spin field exhibits synchronous oscillations, defining the coherent magnetization state. This section
demonstrates the efficiency of the SCHA as a formalism to describe the coherent phase observed in AFMR experiments. 

The depiction of magnetization precession becomes inaccessible when described solely in terms of energy (number) eigenstates. 
The transverse magnetization components, represented by first-order terms of the operators $a$ and $b$, 
result in $\langle S^y\rangle_G=\langle S^z\rangle_G=0$. To appropriately capture the resonating dynamics, the 
Coherent States\cite{rmp62.867} formalism must be employed. Initially applied to describe photons in light 
radiation and HP magnons in FMR experiments\cite{pla29.47,pla29.616,prb4.201}, the CS formalism has more recently 
found application in the SCHA for characterizing spintronics\cite{jmmm472.1,prb106.054313}. In summary, we can define
the coherent state $|\eta\rangle$ as the eigenstate of the annihilation operator, {\it i.e.} 
$\alpha|\eta_\alpha\rangle=\eta_\alpha|\eta_\alpha\rangle$, and similarly for $\beta$ operator. An alternative,
and equivalent, definition is define the coherent state as $|\eta_\alpha\rangle=D(\eta_\alpha)|0\rangle$, where
$D(\eta_\alpha)=\exp(\eta_\alpha \alpha^\dagger-\bar{\eta}_\alpha \alpha)$ is the so-called displacement operator,
and $|0\rangle$ is the vacuum state. 

At finite temperatures, the thermodynamics of coherent states can be derived 
using Thermal Coherent States\cite{pla134.273}. An equivalent and simpler approach to incorporate thermal effects into 
coherent states involves the utilization of the density of states matrix $\rho_{CS}=D(\eta_\alpha)\rho_0 D^\dagger(\eta_\alpha)$,
where $\rho_0=e^{-\beta H_2}/\textrm{Tr}(e^{-\beta H_2})$\cite{jmo38.2339}. The statistical average of an operator $O$ is
then evaluated as $\langle O\rangle=\textrm{Tr}(\rho_{CS}O)$. Employing the displacement operator,
we obtain $\langle \alpha^\dagger \alpha\rangle=|\eta_\alpha|^2+n_\textrm{th}(\epsilon)$, where $|\eta_\alpha|^2$ represents
the condensed part and $n_\textrm{th}(\epsilon)$ is the Bose-Einstein distribution of thermal states with energy $\epsilon$. 
This shift to the CS formalism provides the correct dynamics of coherent magnetization, 
overcoming limitations associated with energy eigenstate descriptions.

The static and oscillating magnetic fields play a central role in generating the CS basis.
Utilizing the bosonic representation, the linear term is written as 
$H_1=\sum_q(h_q^A a_q^\dagger+\bar{h}_q^A a_q+h_q^B b_q^\dagger+\bar{h}_q^B b_q)$,
where
\begin{IEEEeqnarray}{rCl}
	\label{eq.hAhB}
	\IEEEyesnumber
	\IEEEyessubnumber*
	h_q^A(t)&=&\sqrt{\frac{\tilde{S}}{2}}\left[\sqrt{N}(B_E\sin\Delta\theta+\frac{B_D}{2}\sin 2\theta_A)\delta_{q,0}-\right.\nonumber\\
	&&-B_{A,q}^y(t)-iB_{A,q}^z(t)]\gamma\hbar\rho^{1/4}\\
	h_q^B(t)&=&\sqrt{\frac{\tilde{S}}{2}}\left[\sqrt{N}(-B_E\sin\Delta\theta+\frac{B_D}{2}\sin 2\theta_B)\delta_{q,0}-\right.\nonumber\\
	&&-B_{B,q}^y(t)-iB_{B,q}^z(t)]\gamma\hbar\rho^{1/4}.
\end{IEEEeqnarray}
Given that $H_1$ is time-dependent, we employ the Interaction formalism to write the time evolution as 
$O(t)=S^\dagger(t)\hat{O}(t)S(t)$, where the caret denotes time evolution according to the quadratic Hamiltonian $H_2$, and 
$S(t)=T_t\exp[-(i/\hbar)\int_0^t \hat{H}_1(t^\prime)dt^\prime]$ is the S-matrix. In this scenario, 
the time-ordering operator $T_t$ has a minor function and can be replaced by an irrelevant phase. 
In fact, writing $S(t,0)=T_t\prod_i S(t_i+\delta t,t_i)$, where $\delta t\ll 1$, 
and using the Baker-Campbell-Hausdorff Formulae to multiply consecutive time-evolution operators, 
we obtain
\begin{equation}
S(t)=e^{i\vartheta(t)}\exp\left[-\frac{i}{\hbar}\int_{-\infty}^\infty\hat{H}_1(t,t^\prime)dt^\prime\right],
\end{equation}
where $\vartheta(t)$ cancels out when the averages are calculated. Furthermore, we define the retarded Hamiltonian
$\hat{H}_1(t,t^\prime)=\hat{H}_1(t^\prime)\theta(t-t^\prime)$ and adopt an adiabatic time 
evolution initiated from a distant past with the potential turned off, continuing until the present time $t>0$.
To proceed with the evaluation, we must specify the Hamiltonian phase; therefore, let us begin with the AF one.

\textbf{AF phase} - For $\theta_A=0$ and $\theta_B=\pi$, we acquire the 
coefficients $h_q^A(t^\prime,t)=h_q(t^\prime,t)$ and $h_q^B(t^\prime,t)=-\bar{h}_q(t^\prime,t)$, where 
\begin{equation}
h_q(t^\prime,t)=-\sqrt{\frac{\tilde{S}}{2}}\gamma\hbar\rho^{1/4}B_q^{\prime +}(t^\prime)\theta(t^\prime,t)
\end{equation}
with $B_q^{\prime +}=B_q^{y \prime}+iB_q^{z \prime}$. Using Eq. (\ref{eq.abAF}), it is easy to get $S(t)=D(\eta_\alpha)D(\eta_\beta)$, where
\begin{equation}
D(\eta_\alpha)=\prod_q e^{\eta_q^\alpha \alpha_q^\dagger-\bar{\eta}_q^\alpha \alpha_q}
\end{equation}
represent the displacement operator for the $\alpha$-mode, and $D(\eta_\beta)$ is the equivalent expression to the $\beta$-mode.
The coherent state eigenvalues are expressed as
\begin{equation}
\label{eq.etaAFalpha}
\eta_q^\alpha(t)=\sqrt{\frac{\tilde{S}}{2}}\gamma\rho^{1/4}e^{-\Theta_q}\int\frac{d\nu}{2\pi}\frac{\tilde{B}_q^{\prime +}(\nu)e^{i(\omega_q^\alpha-\nu)t}}{\omega_q^\alpha-\nu-i\varepsilon_q},
\end{equation}
and
\begin{equation}
\label{eq.etaAFbeta}
\eta_q^\beta(t)=-\sqrt{\frac{\tilde{S}}{2}}\gamma\rho^{1/4}e^{-\Theta_q}\int\frac{d\nu}{2\pi}\frac{\tilde{B}_q^{\prime -}(\nu)e^{i(\omega_q^\beta-\nu)t}}{\omega_q^\beta-\nu-i\varepsilon_q},
\end{equation}
where we express the magnetic field in frequency space. The parameter $\varepsilon_q\ll \omega_q$ is introduced as an infinitesimal 
value to guarantee convergence in the limit $t\to -\infty$. 
The convergence factor performs the same role as a residual damping term, which measures the dissipation of magnons
in the long-wavelength limit at low temperatures. At $T=0$, the CS ground state is then given by 
$|\eta_q^\alpha\eta_q^\beta\rangle=D(\eta_q^\alpha)D(\eta_q^\beta)|0\rangle$.

When dealing with a monochromatic circularly polarized field at the frequency $\Omega$, the frequency coefficients 
are given by $\tilde{B}_q^{\prime \pm}(\nu)=2\pi B_q\delta(\nu\mp\Omega)$. At the low-temperature limit, thermal fluctuations are 
small and most of the magnons are originating from the coherent behavior. Therefore, employing the monochromatic circularly polarized field, 
we achieve simply results for the population of $\alpha$ and $\beta$ magnons, which are given by
\begin{equation}
N_\alpha=\frac{\sqrt{\rho}\tilde{S}}{2}\sum_qe^{-2\Theta_q}\left|\frac{\gamma B_q}{\Omega-\omega_q^\alpha+i\varepsilon_q}\right|^2
\end{equation}
and
\begin{equation}
N_\beta=\frac{\sqrt{\rho}\tilde{S}}{2}\sum_qe^{-2\Theta_q}\left|\frac{\gamma B_q}{\Omega+\omega_q^\beta+i\varepsilon_q}\right|^2,
\end{equation}
respectively. Note that the momentum distribution is determined by the oscillating field space behavior. Generally, only one well 
defined momentum $q_0$ contributes to the magnon number. In addition, $N_\alpha\gg N_\beta$ ($N_\alpha\ll N_\beta$) for 
$\Omega=\omega_q^\alpha$ ($\Omega=-\omega_q^\beta$), and so only one of the modes is excited. For instance, adopting a 
uniform magnetic field with module $B_\textrm{rf}$ and clockwise orientation, we obtain that the number of magnons is written as
\begin{equation}
N_m^\textrm{AF}=\frac{\sqrt{\rho}}{2}\frac{B_D^r}{\sqrt{2B_E^rB_D^r}}\left(\frac{\gamma B_\textrm{rf}^r}{\varepsilon_0^r}\right)^2 N\tilde{S}
\end{equation}
where we used $e^{-2\Theta_q}=\cosh 2\Theta_q-\sinh 2\Theta_q$ to simplify the expression. 
For \ce{MnF_2}, we use the renormalized parameters $B_E^r=57.1$T and $B_D^r=0.73$T. 
The damping factor is of the order of $\varepsilon_0^r\approx 0.1$ GHz\cite{prb93.014425}, and
we apply a transverse magnetic fields $B_\textrm{rf}^r$ on the order of mT to ensure $\gamma B_\textrm{rf}^r=\varepsilon_0^r$.
Fig. (\ref{fig.Nm} shows the dependence on temperature of $N_m^\textrm{AF}$ and $N_m^\textrm{SF}$ (described below).

\begin{figure}[h]
	\label{fig.Nm}
	\centering 
	\epsfig{file=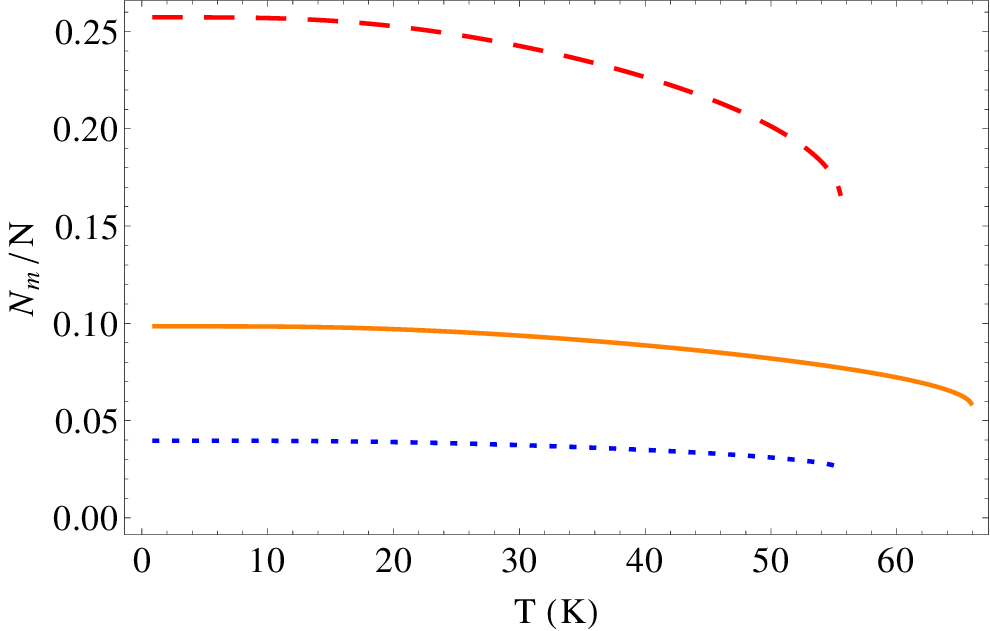,width=0.9\linewidth}
	\caption{(color online) The coherence level $N_m$ for \ce{MnF_2}. The (orange) solid line represents the AF phase, while
	the (red) dashed and (blue) dotted lines define $N_m^\textrm{SF(R)}$ and $N_m^\textrm{SF(L)}$, respectively. 
	Due to the strong magnetic field, the N\'{e}el temperature is reduced to $T=55.5$ K.}
\end{figure}

\textbf{SF phase} - When considering the SF phase, the sublattice angles are given by $\theta_A=-\theta_B=\theta=\arccos[B_x^\prime/(2B_E-B_D)]$,
which reduce the Eqs. (\ref{eq.hAhB}) to
\begin{IEEEeqnarray}{rCl}
	\IEEEyesnumber
	\IEEEyessubnumber*
	h_A(t,t^\prime)&=&-\sqrt{\frac{\tilde{S}}{2}}\gamma\hbar\rho^{1/4}\left[\cos\theta B_q^{y \prime}(t^\prime)+iB_q^{z \prime}(t^\prime)-\right.\nonumber\\
	&&\left.-B_0\right]\theta(t-t^\prime)\\
	h_B(t,t^\prime)&=&-\sqrt{\frac{\tilde{S}}{2}}\gamma\hbar\rho^{1/4}\left[\cos\theta B_q^{y \prime}(t^\prime)+iB_q^{z \prime}(t^\prime)+\right.\nonumber\\
	&&\left.+B_0\right]\theta(t-t^\prime),
\end{IEEEeqnarray}
where $B_0=\sqrt{N}[-(B_E-B_D/2)\sin 2\theta+B_x^\prime\sin\theta]\delta_{q,0}$ is a time-independent parameter associated to uniform fields. 
The determination of the coherent state eigenvalues demand the same procedures of the AF phase. For instance, employing the $\alpha$ and $\beta$ operators, 
we found $S(t)=D(\eta_\alpha)D(\eta_\beta)$, where
\begin{equation}
\eta_q^\alpha(t)=-\sqrt{\tilde{S}}\gamma\rho^{1/4}\frac{e^{-\Xi_q}B_0 e^{i\omega_q^\alpha t}}{\omega_q^\alpha-i\varepsilon_q},
\end{equation}
and
\begin{IEEEeqnarray}{rCl}
\eta_q^\beta(t)&=&\sqrt{\tilde{S}}\gamma\rho^{1/4}\int\frac{d\nu}{2\pi}\left(\frac{e^{\Phi_q}\cos\theta \tilde{B}_q^{y \prime}(\nu)}{\omega_q^\beta-\nu-i\varepsilon_q}+\right.\nonumber\\
&&\left.+\frac{ie^{-\Phi_q}\tilde{B}_q^{z \prime}(\nu)}{\omega_q^\beta-\nu-i\varepsilon_q}\right)e^{i\omega_q^\beta t}.
\end{IEEEeqnarray}
Contrary to the AF phase, the coherence in the SF state manifests distinct characteristics between the $\alpha$ and 
$\beta$ modes. The $\alpha$-mode exclusively involves the time-independent magnetic field component, while the behavior 
of the $\beta$-mode is dynamic, contingent upon the oscillating magnetic field. Additionally, given the uniformity of 
$B_x^\prime$, the $\alpha$-mode solely encompasses the $q=0$ wavenumber, thereby leading to $\Xi_0\to\infty$ (owing to 
$\omega_0^\alpha=0$). Consequently, $|\eta_\alpha\rangle$ is indistinguishable from the vacuum state of the $\alpha$ operators, 
resulting in a lack of coherence within the $\alpha$-mode.  It is important to note that the absence of energy in the 
$\alpha$-mode stems from the $O(2)$ symmetry, which dictates no energy requirement for rotating the spin about the $x$-axis. 
Introducing an easy-plane anisotropy breaks this symmetry, leading to the emergence of a gap, denoted as 
$\Delta_\alpha=\gamma\hbar\sqrt{2B_E B_{Dz}}$, where $B_{Dz}$ represents the field associated with the new easy-plane anisotropy. 
Consequently, a small finite coherence eigenvalue $\eta_\alpha$ arises in this scenario, yet notably distinct from that of 
the $\beta$-mode. Comparable conclusions emerge from a thorough examination utilizing the conventional HP bosonic 
formalism \cite{moura2024}.

In the SF phase, magnons of the $\beta$-mode can couple to photons with either left (clockwise orientation) or 
right (counterclockwise orientation) circular polarization \cite{moura2024}. Adopting a uniform clockwise
circularly polarized magnetic field defined as 
$\bm{B}_\textrm{rf}(t)=B_\textrm{rf}[\cos(\Omega t)\hat{\jmath}^\prime+\sin(\Omega t)\hat{k}^\prime]$, we derive 
$\tilde{B}_q^{y\prime}(\nu)=\pi\sqrt{N}B_\textrm{rf}\delta_{q,0}[\delta(\nu-\Omega)+\delta(\nu+\Omega)]$, 
along with a similar expression for $\tilde{B}_q^{z\prime}(\nu)$. Thus, under the resonating condition $\omega_0^\beta=\Omega$, 
the magnon population is given by
\begin{equation}
N_m^\textrm{SF(L)}=\sqrt{\rho}(e^{\Phi_0}\cos\theta+e^{-\Phi_0})^2\left(\frac{\gamma B_\textrm{rf}^r}{2\varepsilon_0^r}\right)^2 N\tilde{S}.
\end{equation}
The coherence level for clockwise oscillating magnetic fields is obtained through a similar analysis, 
expressed as
\begin{equation}
N_m^\textrm{SF(R)}=\sqrt{\rho}(e^{\Phi_0}\cos\theta-e^{-\Phi_0})^2\left(\frac{\gamma B_\textrm{rf}^r}{2\varepsilon_0^r}\right)^2 N\tilde{S}.
\end{equation}

Notably, the coherence primarily arises from the $\beta$-mode, with the contribution of magnons in the $\alpha$-mode 
amounting to an insignificant fraction on the order of $10^{-8}$ of the total magnons. For the compound \ce{MnF_2} 
($B_E^r=57.1$ T, $B_D^r=0.73$ T, $\varepsilon_0=0.1$ GHz), at $T=10$ K, we derive $N_m^\textrm{SF(L)}/N=0.26$ and 
$N_m^\textrm{SF(R)}/N=0.04$. A comprehensive temperature profile is depicted in Fig. (\ref{fig.Nm}). 
Note that, in the SF phase, the N\`{e}el temperature is reduced to 55.5 K due to the intense magnetic field \cite{pr136.a1068}.
Similar conclusions about the coherence level have been obtained through the application of the HP formalism \cite{moura2024}.

\section{Precession dynamics and susceptibilities}
The magnetic susceptibility is an important quantity in resonance experiments, in which the 
power adsorption provides information about the magnetic damping. Considering only a magnetic microwave field, 
the power absorbed is given by
\begin{equation}
P(t)=\frac{1}{2}\textrm{Re}\int\left(\bm{\bar{H}}\cdot\frac{\partial\bm{B}}{\partial t}\right)dV.
\end{equation}
In momentum space, $\bm{B}_q(t)=\mu_0\bm{H}_q(t)+\int\ud t^\prime \mu_0\bm{\chi}_q(t-t^\prime)\bm{H}_q(t^\prime)$, 
where $\bm{\chi}_q$ is the (retarded) susceptibility tensor. Considering a linearly polarized field $H_q^y$, 
a straightforward evaluation provides 
\begin{equation}
P=\frac{\mu_0\Omega}{2}\sum_q \textrm{Im}\chi_q^{yy}(H_q^y)^2,
\end{equation}
and, for uniform fields, $H_q^y=\sqrt{V}H^y\delta_{q,0}$. The coherent states eigenvalues provide a 
direct evaluation of the magnetic susceptibility. Additionally, in contrast to the LLG approach, which furnishes 
information solely concerning uniform magnetization precession, our findings afford a comprehensive analysis across 
the entire Brillouin zone.

\textbf{AF phase} -  When adopting the coherent behavior, in the low-temperature regime, the dynamics of the transverse components 
can be expressed as
\begin{equation}
\langle S_q^{(y,z)\prime}(t)\rangle=\langle \eta_q^\alpha\eta_q^\beta| \hat{S}_q^{(y,z)\prime}(t)|\eta_q^\alpha\eta_q^\beta\rangle.
\end{equation}
By employing Eqs. (\ref{eq.spinfieldA}) and (\ref{eq.abAF}), we obtain for the $A$ sublattice ($\theta_A=0$)
\begin{IEEEeqnarray}{rCl}
\langle S_{A,q}^+(t)\rangle&=&\sqrt{2\tilde{S}}\rho^{1/4}\left(\eta_q^\alpha\cosh\Theta_q e^{-i\omega_q^\alpha t}+\right.\nonumber\\
&&\left.+\bar{\eta}_q^\beta \sinh\Theta_q e^{i\omega_q^\beta t}\right).
\end{IEEEeqnarray}
One can apply the same procedure for the other sublattice ($\theta_B=\pi$) and obtain
\begin{IEEEeqnarray}{rCl}
\langle S_{B,q}^+(t)\rangle&=&-\sqrt{2\tilde{S}}\rho^{1/4}\left(\eta_q^\alpha\sinh\Theta_q e^{-i\omega_q^\alpha t}+\right.\nonumber\\
&&\left.+\bar{\eta}_q^\beta \cosh\Theta_q e^{i\omega_q^\beta t}\right).
\end{IEEEeqnarray}
After fixing the oscillating mode, the magnetization dynamics derived from SCHA closely mirrors the results 
obtained from HP\cite{jap126.151101}, showcasing the benefits of the renormalization parameter. Unlike the typical 
CS formalism applied to HP bosons, which neglects thermal effects, the SCHA incorporates corrections through the parameter $\rho$.

Note that the transverse spin component depends on the coherent state eigenvalues and, as expected, for vanishing resonance field, 
there is no coherent spin dynamics. Considering the $\alpha$-mode, the ratio of the transverse spin components yields
\begin{equation}
r_q=\left|\frac{\langle S_q^{+A}(t)\rangle}{\langle S_q^{+B}(t)\rangle}\right|=\frac{\omega_q+\gamma[B_D^r+B_E^r(1-\gamma_q)]}{\omega_q-\gamma[B_D^r+B_E^r(1-\gamma_q)]},
\end{equation}
while for the $\beta$-mode, the ratio is the inverse of above equation. For $q=0$ magnons, $r_0=(B_D^r+B_E^r+B_\textrm{sf}^r)/B_E^r\approx 1$,
and both sublattices show counterclockwise precession with similar amplitude in opposite directions, quite different from the behavior 
at the Brillouin zone boundary. When $aq=\pi$, $\langle S_q^{+B}(t)\rangle=0$, which provides dynamics precession only for spins 
on the sublattice A. Additionally, since the oscillating mode is fixed, $|\langle S_{A,q}^y(t)\rangle|=|\langle S_{A,q}^z(t)\rangle|$, 
and similar to the sublattice B. Therefore, magnetization precession trajectory is circular and $\langle S_{A,q}^x\rangle$ is time-independent. 

To calculate the susceptibility, we need to express the spin averages in the laboratory reference frame.
Then, we define the magnetization through $M_q^+(t)=(g\mu_B/a^3)\langle S_q^{\prime+}(t)\rangle$, where 
$\langle S_q^{\prime+}(t)\rangle=\langle S_{A,q}^{\prime+}(t)\rangle+\langle S_{B,q}^{\prime+}(t)\rangle=\sqrt{2\tilde{S}}\rho^{1/4}e^{-\Theta_q}[\eta_q^\alpha(t)e^{-i\omega_q^\alpha t}-\bar{\eta}_q^\beta(t)e^{i\omega_q^\beta t}]$. 
After replacing the coherent state eigenvalues, we perform the Fourier transform to achieve
\begin{equation}
\tilde{M}_q^+=\frac{M_s\gamma e^{-2\Theta_q}(\omega_q^\alpha+\omega_q^\beta)}{(\omega_q^\alpha-i\varepsilon_q-\Omega)(\omega_q^\beta+i\varepsilon_q+\Omega)} \tilde{B}_{rq}^{\prime +},
\end{equation}
where $M_s=g\mu_B \tilde{S}/a^3$ is the saturation magnetization, and the $r$ subindex in $B$ denotes that the field is renormalized as usual. 
Note that, since $\chi\ll 1$, we can write $\bm{B}^\prime\approx\mu_0\bm{H}^\prime$, 
and the magnetization is expressed as $\tilde{M}^+= \chi^+ \tilde{H}_r^{\prime +}$, where the susceptibility is given by
\begin{equation}
\chi_q^+(\Omega)=\frac{2\mu_0M_s\gamma^2 [B_D^r+B_E^r(1-\gamma_q)]}{(\omega_q^\alpha-i\varepsilon_q-\Omega)(\omega_q^\beta+i\varepsilon_q+\Omega)},
\end{equation}
and $\bm{H}_r^{\prime}=\sqrt{\rho}\bm{H}^{\prime}$ is the renormalized H-field (in the laboratory frame reference). 
Since $\chi_q(t)$ is real, $\chi_q^-(\Omega)=\chi_q^+(-\Omega)$. Employing $
\tilde{M}_q^m=\sum_n\chi_q^{mn}\tilde{H}_q^n$, where $\chi_q^{mn}$ is the susceptibility tensor with $\chi_q^{yy}=\chi_q^{zz}=\chi_q^\parallel$ and
$\chi_q^{zy}=-\chi^{yz}=\chi_q^\perp$, it is easy to obtain $\chi_q^+=\chi_q^\parallel+i\chi_q^\perp$. Consequently,
\begin{IEEEeqnarray}{rCl}
\chi_q^{yy}(\Omega)&=&\frac{-2\mu_0M_s\gamma^2[B_D^r+B_E^r(1-\gamma_q)]}{[(\omega_q^\alpha-i\varepsilon_q)^2-\Omega^2][(\omega_q^\beta+i\varepsilon_q)^2-\Omega^2]}[\Omega^2-\nonumber\\
&&-(\omega_q^\alpha-i\varepsilon_q)(\omega_q^\beta+i\varepsilon_q)],
\end{IEEEeqnarray}
and
\begin{IEEEeqnarray}{rCl}
\chi_q^{yz}(\Omega)&=&\frac{-4i\mu_0 M_s\gamma^2[B_D^r+B_E^r(1-\gamma_q)]}{[(\omega_q^\alpha-i\varepsilon_q)^2-\Omega^2][(\omega_q^\beta+i\varepsilon_q)^2-\Omega^2]}(\gamma B_x^{\prime r}-\nonumber\\
&&-i\varepsilon_q)\Omega.
\end{IEEEeqnarray}

Note that the susceptibilities involves only the physical (renormalized) field parameters. The temperature dependence 
arises implicitly in the fields and frequencies. The SCHA provides the complete susceptibility spectrum. 
It can be readily verified that, under the influence of a uniform oscillating field, 
and taking $\varepsilon_q=0$, we retrieve the established results of Keffer and Kittel\cite{pr85.329}, endowed 
the temperature renormalization.

\textbf{SF phase} - Considering that the coherence associated with the $\alpha$ mode is negligible, 
or zero in the absence of easy-plane anisotropy, within the coherent regime, we may employ the effective 
transformation $a_q=b_q=(\cosh\Phi_q \beta_q+\sinh\Phi_q\beta_q^\dagger)/\sqrt{2}$, which results in 
$\langle \bm{S}_{A,q}(t)\rangle=\langle \bm{S}_{B,q}(t)\rangle$. Therefore, we obtain 
$\langle S_q^z(t)\rangle=-\sqrt{\tilde{S}}\rho^{1/4}e^{-\Phi_q}|\eta_q^\beta|\sin(\omega_q^\beta t-\phi_q^\beta)$. 
By remembering that $\varphi$ is a small angle, the usual approximation $S_{(A,B)q}^y\approx \tilde{S}\sqrt{\rho}\varphi_{(A,B)q}$ 
provides $\langle S_q^y(t)\rangle=\sqrt{\tilde{S}}\rho^{1/4}e^{\Phi_q}|\eta_q^\beta| \cos(\omega_q^\beta t-\phi_q^\beta)$.
Then, the transverse spin amplitude ratio gives the ellipticity
\begin{equation}
e_q^\beta=\frac{|\langle S_q^y\rangle|}{|\langle S_q^z\rangle|}=\frac{\sqrt{A+B_q+\epsilon_q^\beta}+\sqrt{A+B_q-\epsilon_q^\beta}}{\sqrt{A+B_q+\epsilon_q^\beta}-\sqrt{A+B_q-\epsilon_q^\beta}}.
\end{equation}
Fig. (\ref{fig.ellipticity}) shows $e_q^\beta$ for the compound \ce{MnF_2} when $B_x^\prime=1.1 B_\textrm{sf}$. 
The ellipticity depends strongly on the wave number and reaches the maximum at the center of the Brillouin zone.
For uniform resonating fields, the magnetization precession is highly elliptical with 
$\langle S_q^y\rangle \approx 25\langle S_q^z\rangle$, which implies a large oscillating magnetization along the 
$x^\prime$-axis. Indeed, due to the elliptical dynamics, it is also possible to induce the precession by
a resonating field applied along the static magnetic field, the so-called parallel pumping process. In this case,
the scattering of a photon of frequency $\omega_p$ results in two magnon with half of the driving frequency\cite{rezende}.
\begin{figure}[h]
	\label{fig.ellipticity}
	\centering 
	\epsfig{file=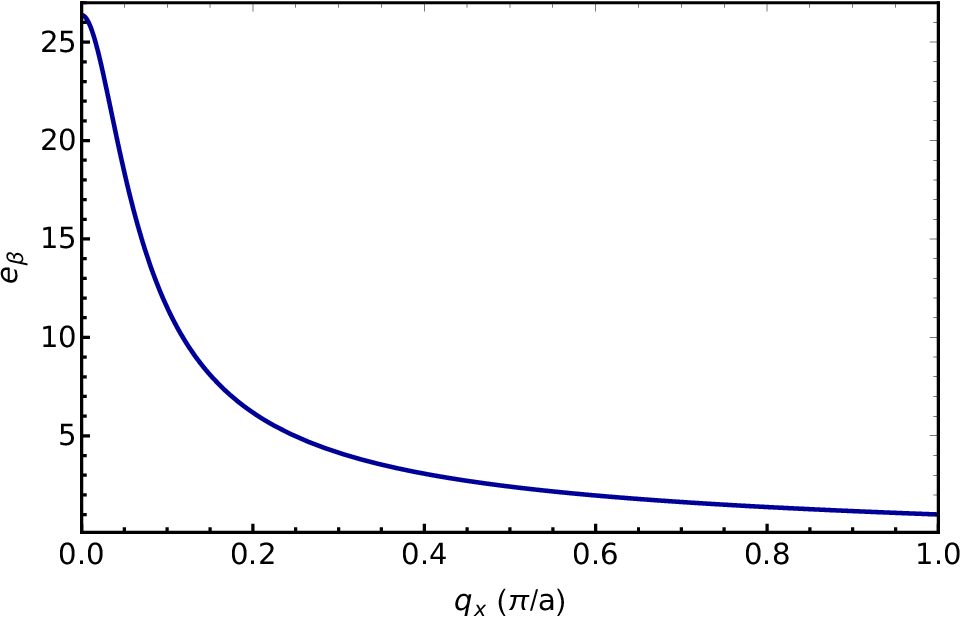,width=0.9\linewidth}
	\caption{The ellipticity of the $\beta$-mode is maximum near the center of the Brillouin zone
	and reaches the minimum at the zone boundary edge.}
\end{figure}

Since the frame was performed along the z-axis, $S_q^{z \prime}=S_q^z$, and the Fourier transform of 
$M_q^z(t)=(g\mu_B/a^3)\langle S_q^{z \prime}(t)\rangle$ results in 
\begin{equation}
\tilde{M}_q^z=2\gamma M_s\frac{i\Omega\cos\theta \tilde{B}_{rq}^{y \prime}-\omega_q^\beta e^{-2\Phi_q}\tilde{B}_{rq}^{z \prime}}{(\Omega+i\varepsilon_q)^2-(\omega_q^\beta)^2}.
\end{equation}
Similarly, $\langle S_q^{y \prime}\rangle=\cos\theta\langle S_q^y\rangle$, and performing the Fourier transform, we obtain
\begin{equation}
\tilde{M}_q^y=-2\gamma M_s\cos\theta\frac{i\Omega \tilde{B}_{rq}^{z \prime}+\cos\theta\omega_q^\beta e^{2\Phi_q}\tilde{B}_{rq}^{y \prime}}{(\Omega+i\varepsilon_q)^2-(\omega_q^\beta)^2}.
\end{equation}
In contrast to the AF scenario, there is no justification for adopting small susceptibility. We employ the relation
$\bm{B}^\prime=\mu_0(\bm{M}+\bm{H}^\prime$ to derive the matrix equation $\Gamma_q^M \tilde{M}_q=\Gamma_q^H\tilde{H}_q$.
The susceptibility exhibits appreciable values only in the resonance vicinity and, for clockwise rotation, we
can approximate $\Omega^2-(\omega_q^\beta)^2$ as $2\Omega(\Omega-\omega_q^\beta)$. Subsequently, the susceptibilities are 
expressed as $\chi_q=(\Gamma_q^M)^{-1}\Gamma_q^H$, yielding the diagonal elements
\begin{equation}
\chi_q^{zz}=-\frac{\omega_se^{-2\Phi_q}}{\Omega-\omega_q^\beta+\omega_s(e^{-2\Phi_q}+\cos^2\theta e^{2\Phi_q})+i\varepsilon_q},
\end{equation}
and $\chi_q^{yy}=\cos^2\theta e^{4\Phi_q}\chi_q^{zz}$. The off-diagonal components are given by
\begin{equation}
\chi_q^{yz}=-\frac{i\omega_s\cos\theta}{\Omega-\omega_q^\beta+\omega_s(e^{-2\Phi_q}+\cos^2\theta e^{2\Phi_q})+i\varepsilon_q},
\end{equation}
while $\chi_q^{zy}=-\chi_q^{yz}$. In the above expressions, we define the angular frequency $\omega_s=\mu_0\gamma M_s$ and
disregarded terms involving $\omega_s^2$ since $\omega_s\ll\omega_q^\beta$. Generally, for $B^{x\prime} \gtrsim B_\textrm{sf}$, 
$\chi_q^{yy}$ is larger than $\chi_q^{zz}$, which justify the elliptical behavior of the magnetization. However, when normalized, both
susceptibilities exhibits the same appearance. Fig. (\ref{fig.chi}) exhibits the diagonal (normalized) 
susceptibility for \ce{MnF_2} considering the static magnetic field as $1.1B_\textrm{sf}$ at $T=10$ K. 
In the SF phase, we observe a behavior similar to the FM resonance. The main change is a slight deviation of 
the resonance point from $\omega_q^\beta$ to $\omega_q^\beta-\omega_s(e^{-2\Phi_q}+\cos^2\theta e^{2\Phi_q})$.
\begin{figure}[h]
	\label{fig.chi}
	\centering 
	\epsfig{file=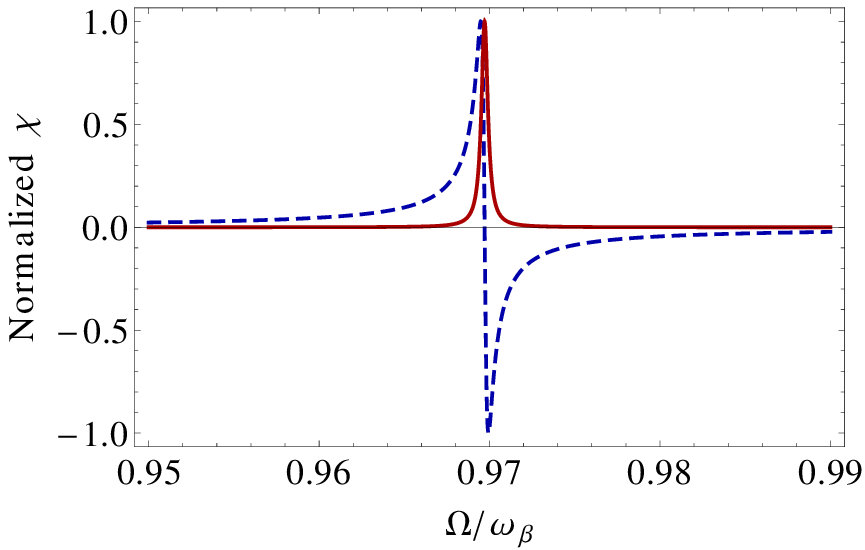,width=0.9\linewidth}
	\caption{(color online) The real (blue dashed line) and imaginary (red solid line) part of the uniform 
	magnetic susceptibility for \ce{MnF_2} when $B^{x\prime}=1.1B_\textrm{sf}$.}
\end{figure}

\section{Conclusions}

This study presented a comprehensive investigation of the AFMR through the utilization of the SCHA. While 
the SCHA has historically been employed to explore phase transitions, its application in spintronic experiments 
involving FM materials has only recently gained attention. However, its applicability to AFM systems remained unexplored.

We conducted an extensive examination of the SCHA, including both semiclassical and quantum approaches, with 
particular emphasis on elucidating the role of the renormalization parameter. Our investigation of the thermodynamic 
properties of both AF and SF phases yields consistent results with those obtained through the phenomenological LLG 
equation and the HP bosonic formalism. Notably, when applied to well-established AFM compounds, the SCHA demonstrated 
remarkable agreement with experimental observations. It is noteworthy that the harmonic approximation facilitated 
temperature corrections more efficiently than the conventional HP formalism, providing superior outcomes.

Furthermore, coherent states were adopted to accurately depict the AFMR dynamics. The magnetization precession dynamics derived 
from the SCHA for the AF phase coincided precisely with those obtained from alternative theoretical methodologies. Conversely, results 
obtained for the SF phase exhibited partial divergence from existing theoretical data. However, in a subsequent submission, we 
investigated the same issue utilizing the HP formalism, obtaining results consistent with those derived from the SCHA. 
Additionally, magnetic susceptibilities were evaluated, yielding congruent outcomes with existing literature, while offering 
the advantage of temperature renormalization.

In conclusion, this study underscores the versatility of the SCHA as an efficient formalism for investigating AFM phenomena in 
spintronics, even close to the temperature transition. Consequently, the method merits consideration as a favorable approach 
for probing numerous spintronic phenomena, particularly those contingent upon thermal effects such as the spin Seebeck effect.

This research was supported by CAPES (Finance Code 001).

\appendix

\section{Renormalization parameter equation}
\label{app.SCHA}
To determine the self-consistent equation for the renormalization parameter, we assess the expected
value of $\langle(\dot{\bm{S}}_q^z)^\dagger\cdot\dot{\bm{S}}_q^z\rangle$, where $(\bm{S}_q^z)^\dagger=(S_{A,q}^z\ S_q^{Bz})$, 
using  both the full Hamiltonian and the quadratic one, ensuring the equality of the outcomes. On one hand, employing the quadratic 
Hamiltonian (\ref{eq.H2}), we obtain
\begin{IEEEeqnarray}{rCl}
\label{eq.SzSzq}
\hbar^2\langle(\dot{\bm{S}}_q^z)^\dagger\cdot\dot{\bm{S}}_q^z\rangle_G&=&\Biggl\langle\frac{\partial H_2}{\partial\varphi_{A,q}}\frac{\partial H_2}{\partial\bar{\varphi}_{A,q}}+\frac{\partial H_2}{\partial\varphi_{B,q}}\frac{\partial H_2}{\partial\bar{\varphi}_{B,q}}\Biggl\rangle_G\nonumber\\
&=&\frac{1}{\beta}(h_{AA,q}^\varphi+h_{BB,q}^\varphi)S^2\rho,
\end{IEEEeqnarray}
with the averages determined using the semiclassical formalism and $\beta=(k_B T)^{-1}$. To ensure Gaussian integration in
$\varphi$ and $S^z$ fields, we extend the limit integration to $-\infty<\varphi,S^z<\infty$. 

On the other hand, for the full Hamiltonian (\ref{eq.Hfull}) we express the same average as
\begin{equation}
\hbar^2\langle(\dot{\bm{S}}_r^z)^\dagger\cdot\dot{\bm{S}}_{r^\prime}\rangle_G=\frac{1}{\beta}\Biggl\langle\frac{\partial^2 H}{\partial\varphi_{A,r}\partial\varphi_{A,r^\prime}}+\frac{\partial^2 H}{\partial\varphi_{B,r}\partial\varphi_{B,r^\prime}}\Biggl\rangle_G,
\end{equation}
where we performed an integration by parts. After differentiation, we take the expected value according to $H_2$, 
causing any odd term in $\varphi$ to vanish. The Fourier transform of the above equation yields
\begin{IEEEeqnarray}{l}
\hbar^2\langle(\dot{\bm{S}}_q^z)^\dagger\cdot\dot{\bm{S}}_q^z\rangle_G=\frac{\gamma\hbar}{\beta S}[-2B_E\cos\Delta\theta\langle f_A f_B\cos\Delta\varphi\rangle_0+\nonumber\\
+B_D(\cos 2\theta_A\langle f_A^2\cos 2\varphi^A\rangle_0+\cos 2\theta_B\langle f_B^2\cos 2\varphi_B\rangle_0)+\nonumber\\
+B_A^x\langle f_A\cos\varphi_A\rangle_0+B_B^x\langle f_B\cos\varphi_B\rangle_0].
\end{IEEEeqnarray}
Given that $\varphi$ and $S^z$ are independent Gaussian distribution, the averages are separable and 
$\langle\cos\varphi\rangle_G=\exp(-\langle\varphi^2\rangle_0/2)$. Therefore
\begin{IEEEeqnarray}{l}
\label{eq.SzSzfull}
\hbar^2\langle(\dot{\bm{S}}_q^z)^\dagger\cdot\dot{\bm{S}}_q^z\rangle_G=\frac{\gamma\hbar S}{\beta}[-2B_E\cos\Delta\theta \rho_E+\nonumber\\
+B_D(\cos 2\theta_A+\cos 2\theta_B)\rho_D+(B_A^x+B_B^x)\rho_Z]
\end{IEEEeqnarray}
where the renormalization equations for $\rho_E$, $\rho_D$, and $\rho_Z$ match those in the main text.
By comparing Eq. (\ref{eq.SzSzq}) with Eq. (\ref{eq.SzSzfull}), we derive the desired equation.

\bibliography{manuscript}

\end{document}